\documentclass[prl,showpacs,twocolumn,typeset,superscriptaddress]{revtex4}

\usepackage{graphicx}
\usepackage{epsfig}
\usepackage{dcolumn}
\usepackage{bm}

\begin{document}

\title{Optimal estimation of quantum processes using incomplete
information: variational quantum process tomography}
\author{Thiago O. Maciel} 
\email{maciel@gmail.com}
\author{Reinaldo O. Vianna}
\email{reinaldo@fisica.ufmg.br}
\affiliation{Departamento de F\'{i}sica - ICEx - Universidade Federal de Minas Gerais,
Av. Ant\^onio Carlos 6627 - Belo Horizonte - MG - Brazil - 31270-901.}

\date{\today}

\begin{abstract}
We develop a quantum process tomography method, 
which variationally reconstruct the map of a process, using noisy and incomplete information about the dynamics.
The new method encompasses the most common quantum process tomography schemes.
It is based on the  variational quantum tomography method (VQT)  proposed by Maciel \emph{et al.} in arXiv:1001.1793[quant-ph]. 
\end{abstract}

\maketitle

The characterization of a quantum system and its dynamics is a daunting challenge. 
The first question which arises in this scenario is what information should we possess
to characterize the dynamics. To answer this question, one needs to choose
a \emph{quantum process tomography} (QPT) scheme. Each procedure demands different resources and operations.

There are four general types of QPT procedures: (i) \emph{standard quantum process tomography} (SQPT)\cite{SQPT}; 
(ii) \emph{ancilla-assisted process tomography} (AAPT)\cite{AAPT1,AAPT2,AAPT3}; 
(iii) \emph{direct characterization of quantum dynamics} (DCQD)\cite{DCQD1,DCQD2,DCQD3};
 (iv) \emph{selective and efficient quantum process tomography}(SEQPT)\cite{SEQTP1,SEQTP2,SEQTP3}. 

In (i) the information is obtained indirectly, performing a set of 
\emph{quantum state tomographies}(QST)\cite{ML1,ML2,ML3,MacielVianna,VQT,Chile}
 of the linear independent states, which spans the Hilbert-Schmidt space of interest, 
after the action of the unknown map.
The second scheme (ii) - also an indirect procedure - makes use of an auxiliary system.
The information is then extracted by means of QST of the joint space (system and ancilla). 
The third one (iii) obtains the dynamical information directly - by means of 
\emph{quantum error detection} (QED)\cite{NielsenChuang} concepts - measuring stabilizers and normalizers.
 Finally, the last method (iv) - which also measures the parameters directly - 
consists in estimating averages over the entire Hilbert space of products of expectation values of two operators. 
For the special case of one-parameter quantum channels, there is also an
interesting method developed by Sarovar and Milburn \cite{Milburn}.

In \cite{VQT},  we developed  a variational  quantum tomography method (VQT), which has the
virtue of yielding a state which is a universal lower bound estimator for every observable. The VQT
approach can reconstruct the state with high fidelity out of incomplete and noisy information.
The method was successfully employed in a quantum optics experiment, where entangled qutrits were
generated \cite{Chile}. In this letter, we extend VQT to the tomography of quantum processes. 
The new method inherits all the advantages of the VQT, and opens the door to the 
characterization of maps in larger systems, where  \emph{myriad} of measurements could be
necessary. With the variational quantum process tomography method (VQPT) we propose here, 
maps can be reconstructed, with high fidelity, using just a fraction of the effort employed
in an Informationally Complete Measurement.

The method we derive has the particular form of  a linear convex optimization problem,
known as Semidefinite Program (SDP),
for which efficient and stable algorithms are available \cite{Boyd, sedumi, yalmip}. 
SDP  consists of minimizing a linear objective under
a linear matrix inequality constraint, precisely,
\begin{center}
{\em minimize} $c^\dagger x$
\begin{equation}\label{sdp}
 subject\,\, to \left\{ F(x)=F_0+\sum_{i=1}^m x_iF_i\geq 0, \right.
\end{equation}
\end{center}
where $c\in C^m$ and  the Hermitian matrices $F_i\in C^{n\times n}$ are given, 
and $x \in C^m$ is the vector of optimization variables.
$F(x)\geq 0$ means that $F(x)$ is a positive matrix.
The problem defined in Eq.\ref{sdp} has no local minima. 
When the unique minimum of this problem cannot be found analytically, 
one can resort to powerful algorithms that return the exact answer \cite{sedumi}.
To solve the problem in Eq.\ref{sdp} could be compared to finding the eigenvalues
of a Hermitian matrix. If the matrix is small enough or has very high symmetry, 
one can easily determine its eigenvalues on the back of an envelope,
but in other cases some numerical algorithm is needed.
Anyway, one never doubts that the eigenvalues of such a matrix can be determined exactly.

A \emph{bona fide} completely positive and non increasing trace map $\mathcal{E}$ can be generally represented as
\begin{equation}\label{cpmap}
\mathcal{E}(\rho) = \sum_{i,j=1}^{d^2}\chi_{ij} E_i \rho E_j^{\dagger},
\end{equation}
where $\rho$ is the system initial state and the \{$E_m$\} form an IC-POVM
(Informationally Complete Positive Operator Valued Measure), 
\emph{i.e.} a complete  basis in the Hilbert-Schmidt space satisfying
\begin{equation}\label{completness}
\sum_{i=1}^{d^2} E_i^{\dagger} E_i = \mathbf{I}.
\end{equation}
 The \{$\chi_{ij}$\} defines the super-operator $\chi$,
 which has all the information about the process.
 It is a Hermitian positive operator.
 Thus the super-operator can be thought as a $d^2 \times d^2 $ density matrix in the Hilbert-Schmidt
 space with $d^4$ independent real parameters (or $d^4-d^2$ in the trace preserving case).

Now we recast both SQPT(i) and AAPT(ii) - which rely on tomography of states -
using the VQT\cite{VQT} methodology. 
We will name the output states of the  unknown map 
 as $ \tilde{\varrho}^k = \mathcal{E}(\rho^k) = \sum_{i,j=1}^{d^2}\chi_{ij} E_i \rho^k E_j^{\dagger}$.
Suppose $n^k$ elements of the POVM ($n^k<d^2$) in the $k^{th}$ output state have been measured, namely
\begin{equation}\label{plamba}
Tr(E_{\lambda}\rho^k) = p_{\lambda}^{k} \mbox{, } \lambda \in [1,n^k].
\end{equation}
Note that $p_{\lambda}^{k}$ are positive numbers, the \emph{known} probabilities.
The probabilities obtained from an experiment are noisy, thus:
\begin{equation}\label{plambadelta}
(1-\Delta_\lambda^k)p_{\lambda}^k \leq Tr(\tilde{\varrho}^k E_{\lambda}^k) \leq (1+\Delta_\lambda^k) p_{\lambda}^k \\ \mbox{, } \lambda \in [1,n^k],
\end{equation}
with $\Delta_\lambda^k$ positive and hopefully small.
Let us refer to  the space spanned by the unmeasured POVM elements as the \emph{unknown} subspace.
Then we can define the \emph{Hamiltonian}
\begin{equation}\label{H}
H^k = \sum_{\lambda=n^k+1}^{d^2}E_{\lambda}^k.
\end{equation}
Thus we obtain a cost function over the unknown subspace, namely,
$\displaystyle Tr(\tilde{\varrho}^k H^k) \equiv Tr(\sum_{i,j=1}^{d^2}\tilde{\chi}_{ij} E_i \rho^k E_j^{\dagger} H^k)$.
This linear functional should be minimized, for we do not know the action of 
the  map on the unknown subspace.

The SQPT method demands quantum state tomography in all linearly independent states 
which span the Hilbert-Schmidt space. With $n^k$ measurements in $k_t$ different states, 
the variational SQPT reads:
\begin{center}
{\em minimize}  $\displaystyle(\sum_k Tr(\tilde{\varrho}^k H^k) + \sum_{\lambda=1}^{n^k} \Delta_\lambda^k)$ 
\begin{equation}
\label{sdp2}
 subject\,\, to
\left\{
\begin{array}{l}
 \tilde{\chi} \geq 0, \\
 Tr(\tilde{\varrho}^k) \le 1,  \\  
\Delta_\lambda^k \geq 0, \\
 (1-\Delta_\lambda^k)p_{\lambda}^k \leq Tr(\tilde{\varrho}^k E_{\lambda}^k) \leq (1+\Delta_\lambda^k) p_{\lambda}^k, \\
 \forall \lambda^k \in [1,n^k]\,\,\,\, and\,\,\,\, k=[1,k_t] . 
\end{array}
\right.
\end{equation}
\end{center}
Eq.\ref{sdp2} returns a map $\tilde{\chi}$ which is the optimal approximation to the
unknown process $\chi$. Note that, at the same time, we were able to identify $\tilde{\varrho}^k$ optimally.
In Fig.1, we illustrate the aplication of the method for the reconstruction of two-qubit
processes.

In the AAPT method, one adds an ancillary system with the same dimension of the main one. 
Then the quantum process takes place  in half subspace, and finally a quantum state tomography is performed
in the whole space, ancilla plus main system. 
The output state now reads 
$ \tilde{\varrho} = (\mathbf{I} \otimes \mathcal{E})(\rho) =
 \sum_{i,j=1}^{d^2}\tilde{\chi}_{ij} (\mathbf{I} \otimes E_i )\rho^k (\mathbf{I} \otimes E_j )^{\dagger}.$ 
With $n$ measurements performed  in this scheme,  AAPT can be recast as
\begin{center}
{\em minimize}  $\displaystyle( Tr(\tilde{\varrho} H) + \sum_{\lambda=1}^{n} \Delta_\lambda)$ 
\begin{equation}
\label{sdp3}
 subject\,\, to
\left\{
\begin{array}{l}
 \tilde{\chi} \geq 0, \\
 Tr(\tilde{\varrho}) \le 1,  \\  
\Delta_\lambda \geq 0, \\
 (1-\Delta_\lambda)p_{\lambda} \leq Tr(\tilde{\varrho} E_{\lambda}) \leq (1+\Delta_\lambda) p_{\lambda}, \\
 \forall \lambda \in [1,n]. 
\end{array}
\right.
\end{equation}
\end{center}

\begin{figure}
\includegraphics[scale=0.41]{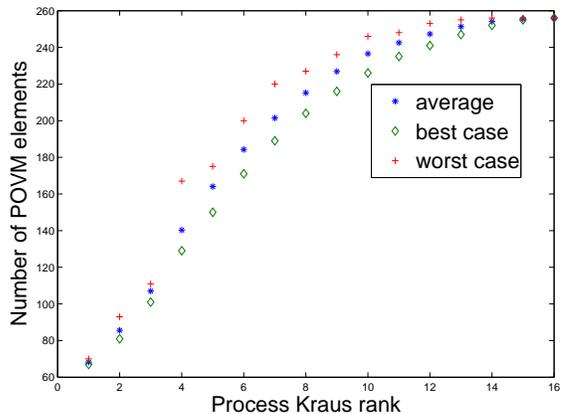}
\caption{Reconstruction of two-qubit processess of all ranks.
We plot the number of independent POVM elements necessary to reconstruct
the map against the process rank. For each rank, we randomly generated 100 processes.}
\end{figure}

In conclusion, we have introduced a new method to perform quantum process tomography, which 
reconstructs a map, with high fidelity, using noisy and incomplete information.
The method is linear and convex, and its unique solution can be obtained very efficiently.
It opens the door to the characterization of the dynamics of larger quantum systems, 
avoiding the need of very large informationally complete measurements.

{\em Acknowledgments}
Financial support by the
Brazilian agencies  FAPEMIG, and  INCT-IQ (National
Institute of Science and Technology for Quantum Information).

\end{document}